\magnification=1200
\vsize = 24truecm
\hsize = 16truecm
\null
\vskip1truecm
\centerline{{\bf MOVING FRAMES HIERARCHY AND BF THEORY}}
\vskip1.5cm
\centerline {Jyh-Hao Lee
\footnote{$^1$}{e-mail:leejh@ccvax.sinica.edu.tw}
and Oktay K. Pashaev
\footnote{$^2$}{e-mail:pashaev@vxjinr.jinr.ru}}
\medskip
\centerline {{\it Institute of Mathematics}}
\centerline {{\it Academia Sinica}}
\centerline{{\it Taipei 11529}}
\centerline {{\it Taiwan}}
\centerline{{\it and}}
\centerline { $^{2}$ {\it Joint Institute for
Nuclear Research}}
\centerline {{\it  Dubna (Moscow) 141980}}
\centerline {{\it Russia}}
\medskip
\vskip1cm
\centerline{{\bf Abstract}}
\baselineskip=16pt
We show that the one-dimensional projection of
Chern-Simons gauged Nonlinear Schr\"odinger model
is eqivalent to an Abelian gauge field theory of
continuum Heisenberg spin chain. In such a theory, the matter field
has geometrical meaning of coordinates in tangent plane to  the
spin phase space, while the $U(1)$ gauge symmetry relates to
rotation in the plane.
This allows us to construct
the infinite hierarchy of integrable gauge theories and
related
magnetic models. To each of them a $U(1)$
invariant gauge fixing constraint of non-Abelian BF
theory is derived.
The corresponding moving frames hierarchy is obtained and the
spectral parameter is interpreted as
a constant-valued statistical gauge potential constrained by
the 1-cocycle condition.
\vskip1cm
\centerline{June 1997}
\vfill\eject
\headline={\centerline {\sevenrm {J.-H.Lee and O.K.Pashaev:
Moving frames hierarchy} }}
\noindent
{\bf I. INTRODUCTION}
\bigskip
\medskip \par
The Nonlinear Schr\"odinger equation (NLS)
in two space dimensions interacting with Abelian
Chern-Simons gauge field (the Jackiw-Pi model)
has attracted much attention recently due to the
beautiful structure of the static limit, admitting N-soliton
solution$^{1}$. After quantization these solitons become
quasiparticles with an arbitrary statistics, called anyons,
while the Chern-Simons gauge field is interpreted
as  the "statistical" one.
The anyons have an interesting
application to the planar physical phenomena as
the Quantum Hall Effect$^{2,3}$.
Very recently attempts to study the generalized statistics
of many particle systems in 1+1 dimensions and the relation with
2+1 dimensional anyons were done$^{4-6}$.
For configurations depending
of the one space direction ( the  {\it lineal} theory), the
reduced theory describes 1+1 dimensional NLS, interacting with an
Abelian BF gauge field. Due to the pure gauge form it is
possible to exclude this field from the gauge invariant description.
However, since in one dimension NLS is an integrable system,
an interesting problem  of the relation between
a long-range structure
of the statistical gauge field in two and integrability in one
dimensions arises.
\par In the present paper we show that the "trivial" gauge field
plays  the crucial role for integrability of the
NLS model. Namely, the homogeneous statistical gauge field
is identical to the spectral parameter for the NLS linear problem.
Our construction is based on a mapping of the reduced Jackiw-Pi model
to non-Abelian BF gauge theory.
This mapping has to appear as the gauge
fixing condition corresponding to the classical Heisenberg spin model.
Then, the matter field has geometrical meaning of
coordinates in tangent to the spin phase space plane, while
$U(1)$ gauge symmetry relates to the rotation in the
plane.
\par In Section I we consider dimensional reduction of the  Jackiw-Pi
model. We find representation of the model
as a $U(1)$ invariant gauge fixing condition of non-Abelian
BF theory. This allows us
to construct an infinite hierarchy of associated integrable models
and the related gauge fixing constraints. In Sec.II we show that the
model has natural interpretation as the gauge theory
of continuum Heisenberg
model defined on the sphere $S^{2}$ or hyperboloid $S^{1,1}$,
according to the sign of nonlinearity. Then, we derive
the associated
Heisenberg and moving frame hierarchies.
Sec.III is devoted to $U(1)$ gauge invariant description of
integrable time hierarchy in the BF theory context. We show
that $B$ field plays the role of squared eigenfunctions
for the Zakharov-Shabat linear problem. In terms of the
recursion operator eigenvalue problem, the hierarchy of constraints
is related to the evolution hierarchy and
higher Poisson structures. In Conclusion we
discuss the chiral solitons for odd members 
of the hierarchy and some open problems.
\bigskip
{\bf II. THE JACKIW-PI MODEL IN 1+1 DIMENSIONS}
\par
\bigskip
The Lagrangian of 2+1 dimensional Nonlinear Schr\"odinger
model interacting with Chern-Simons
gauge field is$^{1}$,
$${\cal L} = {i\over2}(\bar{\psi} D_{0}\psi -
\bar{D}_{0}\bar{\psi}\psi)
-\bar{D}_{1}\bar{\psi} D_{1}\psi
-\bar{D}_{2}\bar{\psi} D_{2}\psi + g|\psi|^{4} +
{k\over4}\epsilon^{\mu\nu\lambda}A_{\mu}\partial_{\nu}
A_{\lambda},     \eqno(2.1)$$
where
$\psi = \psi(x_{1},x_{2},t)$ is complex matter and
$A_{\mu} = A_{\mu}(x_{1},x_{2},t)$, $(\mu = 0,1,2)$,
is a $U(1)$ Abelian (statistical)
gauge field . Here, $D_{\mu} = \partial_{\mu} - {i\over 2}A_{\mu}$,
denotes the
covariant derivative, $g$ and $k$ are the self-interaction and
the Chern-Simons coupling constants respectively.
The related Euler-Lagrange equations of motion are,
$$
iD_{0}\psi + (D^{2}_{1} + D^{2}_{2})\psi + 2g|\psi|^{2}\psi =
0, \eqno(2.2a)$$
$$
\epsilon^{ij}\partial_{i}A_{j} = -{1\over{k}}|\psi|^{2},\eqno(2.2b)$$
$$
\partial_{0}A_{i} - \partial_{i}A_{0} =
-{i\over{k}}\epsilon^{ij}(\bar{D}_{j}\bar{\psi}\psi -
\bar{\psi}D_{j}\psi).
\eqno(2.2c,d)$$
\par  In the special case when the coupling constants
connected by the relation $g = {1\over k}$,
the static configurations
of this model are subject to the self-dual Chern-Simons equations$^{1}$.
The last one admits the linear representation$^{7}$ and has been related
to the Liouville model with  N- vortex/soliton solutions$^{1}$.
However, there is no evidence that
dynamics of these solitons according to eqs.(2.1) is also integrable.
Recently$^{8}$, the integrable dynamics of the Chern-Simons solitons 
has been described
by the Davey-Stewartson equation,
being considered as the 2+1 dimensional extension of the NLS.
Furthermore, it is shown below that model (2.1)
reduced to 1+1 dimension is also
integrable. The  corresponding soliton dynamics subject to the NLS.
\par
Let us consider the boundary as a rectangle on
the plane. Then, the boundary
dynamics of (2.1) is described by the 1+1 reduced model. For
$x_{2}$ independent part
we obtain the Lagrangian,
$${\cal L} = {i\over2}(\bar{\psi} D_{0}\psi -
\bar{D}_{0}\bar{\psi}\psi)
-\bar{D}_{1}\bar{\psi} D_{1}\psi - {1\over 4}B^{2}\bar{\psi}\psi
+ g|\psi|^{4} + {k\over 4}B\epsilon^{\mu\nu}F_{\mu\nu},
\eqno(2.3)$$
where
$B \equiv A_{2}$ is the gauge field
component in the compactified direction,
and $F_{\mu\nu} = \partial_{\mu}A_{\nu}
- \partial_{\nu}A_{\mu}$, $(\mu = 0,1)$.
For the vanishing matter field $\psi = 0$, this Lagrangian reduces
to a pure Abelian $BF$ theory and the model (2.3) can be
considered as the 1+1 dimensional BF gauged NLS. Due to the
non-propagating
character, the gauge field can be eliminated from (2.3).
Nevertheless, as we can see below, (2.3) is equivalent to the
classical spin model and the statistical gauge potential
carries an important information about spectrum of the model.
Moreover, when $\psi \ne 0$, the model (2.3) still is
the BF theory, however for a non-Abelian gauge group in a
special gauge.
\par To proceed first we redefine gauge potentials (and the related
covariant derivatives) as,
$$W_{0} \equiv A_{0} - {1 \over 2} B^{2}\,,\,\, W_{1} \equiv A_{1}.
\eqno(2.4)$$
 Then, the Lagrangian
(2.3) becomes,
$${\cal L} = {i\over2}(\bar{\psi} D_{0}\psi -
\bar{D}_{0}\bar{\psi}\psi)
-\bar{D}_{1}\bar{\psi} D_{1}\psi
+ g|\psi|^{4} + {k\over 4}B\epsilon^{\mu\nu}F_{\mu\nu}
+ \Delta {\cal L}, \eqno(2.5)$$
where,
$$\Delta {\cal L} = - {k \over 6}\partial_{1}B^{3},\eqno(2.6)$$
is the total derivative (which becomes  nontrivial at the
boundaries) and does not influence equations
of motion. The Euler-Lagrange equations for  (2.5) are given by
the system,
$$
iD_{0}\psi + D^{2}_{1}\psi + 2g|\psi|^{2}\psi = 0, \eqno(2.7a)$$
$$\partial_{0}W_{1} - \partial_{1}W_{0} = 0, \eqno(2.7b)$$
and
$$
\partial_{1}B = -{1\over{k}}|\psi|^{2},\eqno(2.8a)$$
$$
\partial_{0}B =
{i\over{k}}(\bar{D}_{1}\bar{\psi}\psi - \bar{\psi}D_{1}\psi),
\eqno(2.8b)$$
where $D_{\mu} = \partial_{\mu} - i/2 W_{\mu}$.
Eqs.(2.7) are exactly gauged NLS model (or the Heisenberg model
in the tangent space form), while (2.8) are the defining relations
for $B$ (the trace of higher dimension) in terms
of the charge conservation law. Hovewer, Eq.(2.8a) - the reduced
Chern-Simons Gauss law  - is more fundamental then (2.8b).
Indeed, Eqs.(2.7) imply the charge
conservation law,
$$\partial _{0}|\psi|^{2} + i\partial _{1}(
(\bar{D}_{1}\bar{\psi}\psi - \bar{\psi}D_{1}\psi) = 0.\eqno(2.9)$$
Then, Eq. (2.8b) arises from the Gauss law (2.8a) and
(2.9) in a similar way to the Chern-Simons theory$^{1}$.
In other words, Eqs.(2.8) guarantee the existence of first
conservation
law for (2.7). In fact, the dual current
$C_{\mu} = \epsilon_{\mu\nu}\partial_{\nu}B$, due to (2.8)
$C_{\mu}$ is conserved.
\par The above systems (2.7) and (2.8) are invariant under $U(1)$
gauge transformations,
$$ \psi \rightarrow \psi e^{i\alpha },
A_{0} \rightarrow A_{0} + 2\partial _{0}\alpha,
A_{1} \rightarrow A_{1} + 2\partial _{1}\alpha,\eqno(2.10) $$
preserving $B$ field: $B \,\rightarrow \, B$.
From Lagrangian (2.5) we recognize that $B$ field plays the role of
Lagrangian multiplier for the zero strength constraint (2.7b).
Moreover, according to the one dimensional Chern-Simons
(or better to say  the BF) Gauss law (2.8a),
creation of
a particle at a point on the line, where $|\psi|^{2} \neq 0$,
is accompanied with creation of the local B field gradient.
This gradient represents one-dimensional analog of the
statistical magnetic field.
For vanishing $\psi$ the
B field is homogeneous, and defined up to the additive constant.
Integrating
(2.8a) along the $x_{1}$ line, then we obtain,
$$
B(+\infty) - B(-\infty) = -{1\over k}\int^{+\infty}_
{-\infty} |\psi|^{2}dx_{1}. \eqno(2.11)$$
For nontrivial charged configurations (solitons) the r.h.s
do not vanish and the
related $B$ field takes the shift on the boundaries.
\par
Actually, Eqs.(2.7) is the 1+1 dimensional analog of
the Ginzburg-Landau equation for superconductor in the
vanishing electric field,
$E = \partial _{0}A_{1} - \partial _{1}A_{0} = 0$.
An important
consequence of the gauge invariance (2.10) is that (2.7) is
independent
of the local gauge transformations parameter $\alpha = \alpha(x,t)$.
This property can be considered
as a generalisation of the well known
isospectrality condition for integrable systems.
Indeed, the second equation (2.7) allows one to exclude the potentials
$A_{0}$ and $A_{1}$ by  $U(1)$ gauge transformation. Due to
(2.7b) a real function $\phi(x,t)$ exists such that
$A_{\mu} = 2\partial _{\mu} \phi $. Then, we define new $\Psi  =
\psi e^{i\phi }$,
subject to  the  Nonlinear
Schr\"odinger equation,
$$
i\partial _{0}\Psi  + \partial ^{2}_{1}\Psi
+ 2g|\Psi |^{2}\Psi  = 0 .
\eqno(2.12)$$
This equation admits an infinite  number
of conservation laws. The conservation law (2.9) in terms
of $\Psi$ is the first member of this
hierarchy, having physical meaning of the charge (number of
particles) conservation.
\par In traditional approach
the NLS integrability follows from
the  Lax  pair or the Zero Curvature representation  with a
constant  spectral parameter, of which the NLS is independent
(isospectrality)$^{9}$. This  fact can be explained now as
the consequence of a $U(1)$ gauge invariance
of the gauged NLS (2.7).
\par The integrability of (2.7) becomes transparent if we
represent the self interaction term in a pure geometrical way.
Thus, for redefined fields,
$$V_{0} = W_{0} + 4g|\psi|^{2}, V_{1}= W_{1},\eqno (2.13)$$
Eqs.(2.7) have the form,
$$
iD_{0}\psi + D^{2}_{1}\psi = 0,\eqno(2.14a)$$
$$
\partial _{0}V_{1} - \partial _{1}V_{0} =
 -4g\partial _{1}|\psi|^{2} ,\eqno (2.14b)$$
of the linear Schr\"odinger equation (the quantum mechanics) in the
electric field proportional to the gradient of the local particles
density. The system (2.14) is equivalent to the following one,
$$
D_{0}\psi = D_{1}\psi_{0},\eqno(2.15a)$$
$$
[D_{0},D_{1}] =
2g(\bar\psi \psi_{0} - \bar\psi_{0}\psi) .\eqno (2.15b)$$
with the constraint,
$$\psi_{0} = iD_{1}\psi.\eqno(2.16)$$
\par The above Eqs. (2.15) define constant curvature
surface with the scalar curvature equal to coupling constant $g$.
In fact, we can show that if $\psi$ and $\psi_{0}$ are the
zweibein fields, then the first equation (2.15a) is the torsionless
condition for definition of the spin connection, identified with the
Abelian gauge field $V_{\mu}$. Then, the the second equation
is just the constant curvature condition,
$$R = g, \eqno(2.17)$$
where R is the scalar curvature, written in terms of zweibeins
and the spin connection. The model (2.17) is known as
the Jackiw-Teitelboim
lineal gravity$^{10}$, while (2.15) as the BF
gauge theoretical formulation of it in terms of the
Einstein-Cartan variables$^{11}$. The system (2.15) appears
as the Euler-Lagrange
equations for the BF action (4.1), discussed in Sec.IV.
Depending on the sign of $g$ the corresponding BF theory has
non-Abelian $SU(2)$,(g = 1), or $SU(1,1)$, (g = -1), gauge
group.
Thus, the reduced
Jackiw-Pi model (2.14) is defined by
gauge condition (2.16) of Euclidean  BF gravity
characterizing  a surface of the Jackiw-Teitelboim model.
\par The mapping (2.15), (2.16) allows us to construct an infinite
hierarchy of gauge constraints, compatible with (2.16),
and describing the hierarchy of corresponding
equations of motion.
Eq.(2.15b)
can be written in the form,
$$\partial_{0}V_{1} - \partial_{1}(V_{0} +
4ig\int^{x}
(\bar\psi\psi_{0} - \bar\psi_{0}\psi)(x')dx') = 0, \eqno(2.18)$$
Then, as in the above procedure from (2.7) to (2.14), but going
in the opposite direction, we introduce the flat Abelian gauge field,
$$ W_{0} = V_{0} +
4ig\int^{x}
(\bar\psi\psi_{0} - \bar\psi_{0}\psi)(x')dx')\,,\,\,
W_{1} = V_{1}. \eqno(2.18) $$
Then, in terms of redefined covariant dirivatives, (2.15)
becomes,
$$ D_{0}\psi = D_{1}\psi_{0} + 2g\psi
\int^{x}
(\bar\psi\psi_{0} - \bar\psi_{0}\psi)(x')dx')\,,\,\,
 \eqno(2.19a) $$
$$\partial_{0}W_{1} - \partial_{1}W_{0} = 0. \eqno(2.19b)$$
Of course,
for nonsingular gauge configurations
we can always exclude
field $A_{\mu}$. But, we find it is important to keep
this gauge field, playing the role similar to
the spectral parameter in the usual Inverse Scattering approach.
Then, model (2.12) written in terms of
$\Psi$, is explicitly invariant under the local gauge transformations
(2.10) and the gauge
invariance plays the role of  generalized
isospectrality.
Below we show how far this analogy can be continued.
It is convenient to write Eq.(2.19a) and its complex conjugate
in the matrix form,
$$
\left(\matrix{D_{0}\psi \cr \bar D_{0}\bar \psi  \cr}
\right) =
\left(\matrix{
D_{1} + 2g\psi\int^{x}\bar\psi & -2g\psi\int^{x}\psi  \cr
-2g\bar\psi\int^{x}\bar\psi & \bar D_{1} + 2g\bar\psi
\int^{x}\psi   \cr}
\right)
\left(\matrix{ \psi_{0} \cr \bar \psi_{0}  \cr }
\right),\eqno(2.20)$$
Then, the constraint (2.16) can be written as,
$$
\left(\matrix{ \psi^{(1)}_{0}\cr \bar\psi^{(1)}_{0} \cr} \right)
= i\sigma_{3}
\left(\matrix{ D_{1}\psi \cr \bar D_{1}\bar\psi \cr} \right)
= \Lambda
\left(\matrix{ \psi\cr \bar\psi \cr} \right)\eqno(2.21)$$
where we introduced the integro-differential opreator
$$\Lambda =
i\sigma_{3}
\left(\matrix{
D_{1} + 2g\psi\int^{x}\bar\psi & -2g\psi\int^{x}\psi  \cr
-2g\bar\psi\int^{x}\bar\psi & \bar D_{1} + 2g\bar\psi
\int^{x}\psi   \cr}
\right).\eqno(2.22)$$
For skew-symmetric linear operator (see Sec.IV) we have
to consider the
integral part defined by the symmetric boundaries
$$\int^{x} f(y)dy = {1 \over 2}(\int^{x}_{-\infty}f(y)dy
+ \int^{x}_{\infty} f(y)dy).\eqno(2.22a)  $$
The above operator $\Lambda$ represents the $U(1)$
gauge covariant
extension of the AKNS operator ${\cal L}$ $^{12}$.
In fact,
$$\Lambda = {\cal L} + {1 \over 2}W_{1}I,\eqno(2.23a)$$
where
$${\cal L} =
i\sigma_{3}
\left(\matrix{
\partial_{1} + 2g\psi\int^{x}\bar\psi & -2g\psi\int^{x}\psi  \cr
-2g\bar\psi\int^{x}\bar\psi & \partial_{1} + 2g\bar\psi
\int^{x}\psi   \cr}
\right).\eqno(2.23b)$$
Here
$I$ is the identity matrix, and $\Lambda$ is covariant
under $U(1)$ gauge transformations (2.10).
Then, gauged NLS (2.7) has the form,
$$
i\sigma_{3}\left(\matrix{ D_{0}\psi \cr
\bar D_{0}\bar\psi \cr} \right)
= \Lambda^{2}
\left(\matrix{ \psi\cr \bar\psi \cr} \right).\eqno(2.24)$$
This representation suggests how to create the whole hierarchy
of equations associated with gauged NLS (2.24).
We define the set of constraints,
$$
\left(\matrix{ \psi^{(n)}_{0} \cr
\bar \psi^{(n)}_{0} \cr} \right)
= \Lambda^{n}
\left(\matrix{ \psi\cr \bar\psi \cr} \right),\eqno(2.25)$$
labeled with an integer $n$.
As mentioned above, (2.15) describes a classical
motion for the BF theory, with $\psi_{0}$ playing the role of
Lagrange multipliers. The arbitrariness of $\psi_{0}$ guarantees the
general time reparametrization invariance of the theory, while a
specific choice for $\psi_{0}$ defines the
corresponding evolution. The hierarchy (2.25) can be considered as
the hierarchy of gauge fixing constraints of BF theory.
Then, every
constraint is related to the nonlinear spin model (Heisenberg
hierarchy), being just the tangent  space representation for the
model.  \par
The first few constraints from (2.25) are given by,
$$
\left(\matrix{ \psi^{(0)}_{0} \cr
\bar \psi^{(0)}_{0} \cr} \right)
=
\left(\matrix{ \psi\cr \bar\psi \cr} \right),\eqno(2.26)$$
$$
\left(\matrix{ \psi^{(1)}_{0} \cr
\bar \psi^{(1)}_{0} \cr} \right)
= i\sigma_{3}
\left(\matrix{ D_{1}\psi\cr \bar D_{1}\bar\psi \cr} \right),
\eqno(2.27)$$
$$
\left(\matrix{ \psi^{(2)}_{0} \cr
\bar \psi^{(2)}_{0} \cr} \right)
= -
\left(\matrix{ D^{2}_{1}\psi + 2g|\psi|^{2}\psi
\cr \bar D^{2}_{1}\bar\psi + 2g|\psi|^{2}\bar \psi
\cr} \right),\eqno(2.28)$$
$$
\left(\matrix{ \psi^{(3)}_{0} \cr
\bar \psi^{(3)}_{0} \cr} \right)
= - i\sigma_{3}
\left(\matrix{ D^{3}_{1}\psi + 6g|\psi|^{2}D_{1}\psi
\cr \bar D^{3}_{1}\bar\psi + 6g|\psi|^{2}\bar D_{1}\bar \psi
\cr} \right).\eqno(2.29)$$
\par Thus, for every $n$, the corresponding $\psi^{(n)}_{0}$
defines the evolution
equation,
$$
i\sigma_{3}\left(\matrix{ D_{0_{n}}\psi \cr
\bar D_{0_{n}}\bar\psi \cr} \right)
= \Lambda^{n}
\left(\matrix{ \psi\cr \bar\psi \cr} \right)
= \left(\matrix{\psi^{(n)}_{0} \cr
\bar\psi^{(n)}_{0}  \cr} \right),\eqno(2.30a)$$
$$[D_{0_{n}}, D_{1}] = 0.
\eqno(2.30b)$$
The first members of the hierarchy, (n = 1, 2, 3), are,
$$D_{0_{1}}\psi - D_{1}\psi = 0, \eqno(2.31)$$
$$iD_{0_{2}}\psi + D^{2}_{1}\psi + 2g|\psi|^{2}\psi = 0,\eqno(2.32)$$
$$D_{0_{3}}\psi + D^{3}_{1}\psi + 6g|\psi|^{2} D_{1}\psi = 0.
\eqno(2.33)$$
This $U(1)$ gauged hierarchy has the flat Abelian connection (2.30b).
It means that every equation (2.30) in terms of the gauge
invariant variables, likes (2.12), reduces to the form of the
usual NLS hierarchy. Indeed, due to (2.30b) for any $n$
there exists a real function
$\alpha_{n} = \alpha_{n}(x,t_{n})$ such that,
$$W_{1} = \partial_{1}\alpha_{n},\,\, W_{0_{n}} = \partial_{0_{n}}
\alpha_{n}.\eqno(2.34)$$
Therefore, for the gauge invariant fields $\Psi =
\psi e^{i\alpha_{n}}$ the hierarchy (2.30) reduces to the NLS one,
$$
i\sigma_{3}\left(\matrix{ \partial_{0_{n}}\Psi \cr
\partial_{0_{n}}\bar\Psi \cr} \right)
= {\cal L}^{n}
\left(\matrix{ \Psi\cr \bar\Psi \cr} \right)
= \left(\matrix{\Psi^{(n)}_{0} \cr
\bar\Psi^{(n)}_{0}  \cr} \right).\eqno(2.35)$$
However, the gauged hierarchy (2.30) contains much more
information in the addition. As we show in the next section,
restricted on the subclass of  constant gauge potentials,
Eqs.(2.30) provide
the linear problem for any of Eqs.(2.35).
Moreover, every equation of the hierarchy (2.30)
can be reduced dimensionally from the related 2+1 dimensional
Chern - Simons gauged theory.
\bigskip
\medskip
\noindent 
{\bf III. ABELIAN GAUGE  THEORY FOR MOVING FRAMES HIERARCHY}
\par
\bigskip
\par
The hierarchy of Abelian gauge theories described in the previous
section by (2.30), is equivalent to the system (2.15) with the
hierarchy of U(1) gauge invariant constraints (2.25). But system
(2.15) is the constant curvature surface equation written in the
Einstein-Cartan zweibein formalism.
As we see below, every constraint (2.25) supplied to
(2.15), defines an evolution equation for three dimensional
unit vector ${\bf s}$, and can be considered as nonlinear
$\sigma$ model on the sphere $S^{2}$ or pseudosphere $S^{1,1}$.
\par
Let us consider the Lie group $G$ with element $g$, generated by
$\tau_{i} (i=1,2,3)$ , satisfying to the relations,
$$
\tau_{i}\tau_{j} = h_{ij} + ic_{ijk}\tau_{k},
\eqno(3.1)$$
where $h_{ij}$ and $c_{ijk}$ are the Killing metric
and the structure constants of the algebra respectively.
We define an orthonormal trihedral set of unit vectors
${\bf n}_{i}$
in the adjoint representation,
$$
({\bf n}_{i},\tau) =
{\bf n}^{k}_{i}\tau_{k} = h_{kl}{\bf n}^{k}_{i}\tau^{l} =
g \tau_{i} g^{-1},\eqno (3.2)$$
The Killing metric $h_{ij}$ and structure constants
$c_{ijk} = - c_{jik}$ defines correspondingly the inner and the
cross product between
three-vectors, transforming in the adjoint representation:
$$
({\bf n}_{i},{\bf n}_{j}) = h_{ij} ,\eqno (3.3a)$$
$$
{\bf n}_{i}\wedge  {\bf n}_{j} = c_{ijk} {\bf n}_{k}.\eqno (3.3b)$$
\par
Given smooth vector fields ${\bf n}_{i}= {\bf n}_{i}(x,t)$
define at each space $x$ and time $t$
three vectors $({\bf n}_{1}(x,t),{\bf n}_{2}(x,t),{\bf n}_{3}(x,t))$
forming an orthonormal
basis, called the moving frame.
The chiral current,
$$
J_{\mu } = g^{-1}\partial _{\mu }g,  \eqno(3.4)$$
in the adjoint representation defines rotation of the
moving frame by equations,
$$
\partial _{\mu }{\bf n}_{i}= (J^{ad}_{\mu })_{ik}{\bf n}_{k},
\eqno(3.5)$$
where,
$$
(J^{ad}_{\mu })_{ik} = - ic_{ijk}(J_{\mu })_{j} =
 i(J_{\mu })_{j}c_{jik} ,
\eqno(3.6)$$
and $J_{\mu } = \sum  (J_{\mu })_{j} (1/2)\tau_{j}$ .
Matrices $J^{ad}_{\mu }$ have the symmetry,
$(J^{ad}_{\mu })_{ij} h_{jj} = - (J^{ad}_{\mu })_{ji} h_{ii}$,
and are antisymmetric for $SU(2)$ case $h_{ij} = \delta _{ij}$
, $c_{ijk} =
\epsilon _{ijk}$.
\par
The current (3.4) satisfies the zero curvature equations,
$$
\partial _{\mu }J_{\nu } -
\partial _{\nu }J_{\mu } + [J_{\mu }, J_{\nu }] = 0 ,
\eqno(3.7)$$
as compatibility conditions for the system (3.5).
We decompose  matrix $J_{\mu }$ to the diagonal and off diagonal
parts,
$J_{\mu } = J^{(0)}_{\mu }+ J^{(1)}_{\mu },$
parametrized in the form$^{13}$,
$$
J^{(0)}_{\mu } = i/4 \sigma _{3}V_{\mu }\,\, ,
\,\,\,J^{(1)}_{\mu } =
\left(\matrix{0&-g\bar \psi_{\mu} \cr \psi_{\mu}&0 \cr}\right),
\eqno(3.8)$$
where $g = + 1$ for $su(2)$, and $g = - 1$,
for $su(1,1)$ cases. Then,  in  the
adjoint representation (3.6) we have the form,
$$
(J_{\mu })^{ad} = {1\over 2}
\left(\matrix{
0&V_{\mu}&4g\Re (\psi_{\mu})\cr
-V_{\mu}&0&4g\Im (\psi_{\mu}) \cr
-4\Re (\psi_{\mu})&-4\Im (\psi_{\mu})&0 \cr}\right).
\eqno(3.9)$$
\par The moving frame rotates according to equations,
$$
\partial _{\mu }{\bf n}_{i}= -1/2 V_{\mu }
\epsilon _{ij}{\bf n}_{j} - g U_{i\mu }{\bf s} ,\eqno(3.10a)
$$
$$
\partial _{\mu }{\bf s} = U_{i\mu }{\bf n}_{i},
\eqno(3.10b)$$
where $U_{\mu } \equiv 2(\Re (\psi _{\mu}),\Im (\psi _{\mu }))$.
For vector ${\bf s \equiv  n}_{3}$ the constraint,
$({\bf s}(x,t), {\bf s}(x,t)) = h_{33} ,$
is valid, where $h_{33}= 1$,
which means that it belongs to two-dimensional sphere $S^{2}$ or
pseudosphere $S^{1,1}$,
correspondingly.
Two vector fields $({\bf n}_{1}(x),{\bf n}_{2}(x))$ at each $(x,t)$
form a
basis in the tangent plane to  the corresponding  manifold
for ${\bf s}(x)$.
However, by eq.(2.3)
vectors ${\bf n}_{1}$ and ${\bf n}_{2}$ are not
determined uniquely.
If  we choose the other pair ${\bf n}_{1}' , {\bf n}_{2}'$, as the
rotated basis,
$${\bf n}_{1}'= \cos  \alpha  {\bf n}_{1} -
\sin  \alpha  {\bf n}_{2} , {\bf n}_{2}'= \cos
\alpha  {\bf n}_{2} + \sin  \alpha  {\bf n}_{1} , \eqno(3.11)$$
the related $V_{\mu }'$ and $\psi_{\mu }'$
are the $U(1)$ gauge
transformed fields,
$$
V_{\mu }'= V_{\mu } + 2 \partial _{\mu }\alpha  ,
\psi_{\mu }'= e^{i\alpha }\psi_{\mu }
\eqno(3.12)$$
The expression (3.11)  suggests us to
introduce a complex basis
${\bf n}_{+}= {\bf n}_{1} + i {\bf n}_{2} , {\bf n}_{-}=
{\bf n}_{1} - i {\bf n}_{2} ,$
satisfying the following relations,
$$
({\bf n}_{\pm}, {\bf n}_{\pm}) = 0\,\,,\,\,
({\bf n}_{+}, {\bf n}_{-}) = 2g,\eqno(3.12a)
$$
$$
{\bf n}_{+}\times  {\bf s} = i{\bf n}_{+} ,
{\bf n}_{-}\times  {\bf s} = -i{\bf n}_{-} ,
{\bf n}_{-}\times  {\bf n}_{+} = 2 i g{\bf s}. \eqno(3.12b)
$$
Then, we get,
$$
\psi_{\mu } = 1/2 \kappa^{2}(\partial _{\mu }{\bf s}, {\bf n}_{+}),
\bar{\psi}_{\mu } =
 1/2 \kappa^{2}(\partial _{\mu }{\bf s}, {\bf n}_{-}) .
\eqno(3.13)$$
 In terms of (3.12) the moving frame system (3.10) becomes,
$$
D_{\mu }{\bf n}_{+} = -2 g\psi_{\mu } {\bf s} ,\eqno(3.13a)
$$
$$
\partial _{\mu }{\bf s} = \psi_{\mu } {\bf n}_{-} +
\bar{\psi}_{\mu } {\bf n}_{+} ,\eqno (3.13b)
$$
where $D_{\mu } \equiv  \partial _{\mu } - i/2 V_{\mu }$,
is $U(1)$ covariant derivative.
This   form   is   explicitly invariant under the local $U(1)$
gauge transformations
$$
{\bf n}_{+}\rightarrow e^{i\alpha }
{\bf n}_{+} \,,\, {\bf n}_{-} \rightarrow e^{-i\alpha } {\bf n}_{-}\, ,\,
V_{\mu} \rightarrow V_{\mu} + 2\partial_{\mu}\alpha \,,\,
{\bf s} \rightarrow {\bf s} \,,
\eqno(3.14)$$
 that are just the local rotations in tangent to the
vector {\bf s} plane.
From (3.13) follows that $V_{\mu }$ and $\psi_{\mu }$ fields
subject to the system (the integrability condition),
$$
D_{0}\psi = D_{1}\psi_{0}\qquad , \eqno(3.15a)
$$
$$
[D_{0},D_{1}] = 2g(\bar{\psi}\psi_{0} -
\bar{\psi}_{0}\psi), \eqno(3.15b)
$$
where we skip the index for $\psi_{1}$ field.
This system coincides with (2.15). Moreover, the
reduced Jackiw-Pi model (2.7) corresponds to the constraint (2.16).
Under this constraint, the moving frame evolution (3.13) describes,
$$
D_{0}{\bf n}_{+} = - 2giD_{1}\psi {\bf s}
,\,\,
D_{1}{\bf n}_{+} = - 2g\psi{\bf s} ,\eqno (3.16a,b)
$$
$$\partial _{0}{\bf s} = iD_{1}\psi {\bf n}_{-} - i\bar D_{1}\bar \psi
{\bf n}_{+}\,\,,\,\,
\partial _{1}{\bf s} = \psi {\bf n}_{-} + \bar \psi {\bf n}_{+}
,\eqno (3.16c,d)$$
Differentiating (3.16d),
$$
\partial ^{2}_{1}{\bf s} = D_{1}\psi{\bf n}_{-}
+ \bar{D}_{1}\bar \psi {\bf n}_{+} - 4g|q_{1}|^{2} {\bf s},
\eqno (3.17)
$$
we immediately see that vector ${\bf s}$ satisfies the
Landau-Lifshitz
equation for continuum isotropical Heisenberg spin chain model,
$$
\partial _{0} {\bf s} =
{\bf s \times  \partial }^{2}_{1}{\bf s},
\eqno(3.18))$$
where ${\bf s}$ belongs to the 2-dimensional
sphere $S^{2}$ (g = 1)$^{14}$, or  pseudosphere
$S^{1,1}$ (g = -1)$^{15}$.
The above results illuminate the role of the Chern-Simons
statistical gauge field
in the Heisenberg model. Two components, $A_{0}, A_{1}$ are just the
gauge degrees of freedom of rotation in the tangent plane,
while the $B = A_{2}$ component is defined by
relations
$$\partial_{1} B = - {1 \over 4gk}(\partial_{1}{\bf s})^{2},
\eqno(3.19a)$$
$$\partial_{0} B = {1\over k}\partial_{1}{\bf s}(\partial^{2}_{1}{\bf s}
\wedge {\bf s}),\eqno(3.19b)$$
The right hand side of Eq.(3.19a) is the energy density of the
model (3.18). As in Sec.II, (see Eq.(2.8a)), the local magnetic
energy is always accompanied with the gradient of the $B$ field.
From above Eqs.(3.19) it is evident that the total magnetic energy is
conserved quantity,  defining the
asymptotic jump
$$
B(+\infty) - B(-\infty) = -{1\over 4gk}\int^{+\infty}_
{-\infty} (\partial_{1}{\bf s})^{2}dx. \eqno(3.20)$$

\par As shown in Sec.II, the gauge constraint (2.16) is
the second member of the infinite hierarchy of the gauge
constraints (2.25). To show that to every constraint from
this hierarchy corresponds the moving frame system and the
spin model we can proceed analogously to the case (2.18-19).
After redefinition (2.18) the system (3.13)
becomes,
$$
\left(\matrix{D_{0}{\bf n}_{+} \cr \bar D_{0}{\bf n}_{-} \cr}
\right) = - 2g {\bf s}
\left(\matrix{\psi_{0}  \cr
\bar\psi_{0}   \cr}
\right) + 2g
\left(\matrix{ {\bf n}_{+} \cr -{\bf n}_{-}  \cr }
\right) \int^{x}(\bar\psi\psi_{0} - \bar\psi_{0}\psi)
,\eqno(3.21)$$

$$
\left(\matrix{D_{1}{\bf n}_{+} \cr \bar D_{1}{\bf n}_{-} \cr}
\right) = - 2g {\bf s}
\left(\matrix{\psi  \cr
\bar\psi   \cr}
\right) ,\eqno(2.22)$$

$$\partial_{0}{\bf s} = \psi_{0}{\bf n}_{-} + \bar\psi_{0}{\bf n}_{+}
\,\,,\,\,
\partial_{1}{\bf s} = \psi{\bf n}_{-} + \bar\psi{\bf n}_{+}.
\eqno(2.23)$$
The tangent space vectors evolution (3.21) is convenient to
be rewritten as,
$$
\left(\matrix{D_{0}{\bf n}_{+} \cr \bar D_{0}{\bf n}_{-}  \cr}
\right) = -2g
\left(\matrix{
{\bf s} - {\bf n}_{+}\int^{x}\bar\psi & {\bf n}_{+}\int^{x}\psi  \cr
{\bf n}_{-}\int^{x}\bar\psi & {\bf s} - {\bf n}-{-}
\int^{x}\psi   \cr}
\right)
\left(\matrix{ \psi_{0} \cr \bar \psi_{0}  \cr }
\right),\eqno(3.24)$$
Then, we note that compatibility condition for system (3.22-24)
is given by (2.20). Moreover, the hierarchy of the gauge fixing
constraints (2.25) produces the hierarchy of the moving frame
evolutions (3.24) and the related higher order analogs of the NLS
(2.30).
This moving frame hierarchy generates also the higher order
analogs of the Heisenberg model (3.18). To follow this direction
we need to extract information only in terms
of the spin vector ${\bf s}$. First we note, that
the integrand of (2.21)
$$2ig(\bar\psi\psi_{0} - \bar\psi_{0}\psi) = {\bf s}(\partial_{0}
{\bf s}
\wedge\partial_{1}{\bf s}) ,\eqno(3.25)$$
is the topological charge density for the spin configuration
on the space-time worldsheet, or the volume element on the spin
phase space. Then, we obtain the recursion relations for the
evolution equations,
$$\partial_{0_{n}}{\bf s} = [{\bf s}\wedge \partial_{1}
- \partial_{1}{\bf s}\int^{x} {\bf s}(\partial_{1}{\bf s}\wedge .
)]
\partial_{0_{n-1}}{\bf s}, \eqno(3.26)$$
where integer $n$ describes the n-th member of the hierarchy.
The first few evolutions are given by,
$$\partial_{0}{\bf s} = \partial_{1}{\bf s},\eqno(3.27)$$
$$\partial_{0}{\bf s} = {\bf s}\wedge \partial^{2}_{1}{\bf s},
\eqno(3.28)$$
$$\partial_{0}{\bf s} =
{\bf s}\wedge{\bf s}\wedge\partial^{3}_{1}{\bf s}
- {3\over 2}(\partial_{1}{\bf s}\partial_{1}{\bf s})
\partial_{1}{\bf s}.
\eqno(3.29)
$$
\par The moving frame approach allows us to
introduce also the traditional
zero curvature description. To proceed, we need to find an
appropriate
reduction for (3.7-8).
We note that the NLS (2.12) can be considered as a quantum
mechanics with the potential function proportional to
the probability density,
$U(x) = \bar\Psi(x)\Psi(x)$. But, as well known$^{16}$, the
Galilean boosts representation in the quantum mechanics makes use of
1-cocycle $\omega_{1}$:
$$ U(v)\Psi (x) = e^{-2\pi i \omega _{1}(x;v)}\Psi (x - vt),
\eqno(3.30)$$
with the following condition to satisfy,
$$\omega _{1}(x - ut;v) - \omega _{1}(x;u + v) +
\omega _{1}(x;u) = 0,
(mod \, Z),\eqno(3.31)$$
where $U(v)$ is the unitary operator shifting the coordinates
as
$$U(v)xU^{-1}(v) = x - vt.\eqno(3.32)$$
Then, condition (3.31) ensures the group multiplication for $U(v)$
operators:
$$U(v)U(u) = U(u + v).\eqno(3.33)$$
The quantity $\omega_{1}(x;v)$ is given by
$$ 2\pi\omega _{1}(x;v) = -{v^{2}\over 4}t + {v\over 2}x ,
\eqno(3.34)$$
and provides the local phase transformation for the wave function.
Thus, the NLS model (2.12)
is invariant under the Galileo transformations
$$x \rightarrow x'= x - vt,\,\, t \rightarrow t' = t,\eqno(3.35a) $$
$$\Psi \rightarrow \Psi'\exp {i({v^{2}\over 4}t'
+ {v\over 2}x')}.\eqno(3.35b)$$
This invariance can be interpreted in two ways. First, that the
space-time transformation to constant velocity moving frame should
be accompanied with the phase transformation for the "wave function"
$\Psi$. This is the 1-cocycle representation as given above.
 In the second interpretation we can say, that
any one-parameter group ($v$ = const) of the phase
transformations (3.35b)
becomes the symmetry of the model (2.12) if we transform
coordinates to
the moving frame with velocity $v$ according to (3.35a).
The second
interpretation has the flavor of the U(1) Abelian local gauge theory
with the gauge potentials transforming on the constant value.
In fact, the
linear in space and time
phase transformations (3.35b) satisfying (3.31)
generate a subgroup of the Abelian gauge transformations, shifting
$W_{1}$ potential (2.4) on the constant velocity.
This is why it is of interest to consider the restricted class of the
constant gauge potentials. As shown below they are associated
with the spectral parameter in the linear problem for the
NLS. Therefore, in our approach the whole structure of the linear
problem with the spectral parameter has the origin from the $U(1)$
gauge potential $W_{1}$, while the related phase is the linear
function of space and time.
\par
Our idea has been suggested by some known facts from the
theory of
super conductivity $^{17}$ and the field theory with
curved momentum
space$^{18}$. For the superconductor in the ground state
under the external
electromagnetic action the generalised canonical momentum
${\bf p} = m{\bf v} + e/c {\bf A}$ vanishes.
This fact
provides the relation between superconducting electrons velocity
and the vector potential,
$$ {\bf v} = - e/c {\bf A},\eqno(3.36) $$
restricting Abelian gauge transformations to the London gauge $^{17}$,
$$ div {\bf A} = 0.\eqno(3.37)$$
In one space dimension this gauge leads to
the  constant  gauge potential $A$.
Furthermore, for constant potentials
$A_{\mu}$, transition to the new momentum,
$$p_{\mu} \rightarrow p_{\mu} - c_{\mu},\eqno(3.38)$$
generates the group of parallel transitions in
the momentum space. For the non - relativistic theory it is just
the Galileo transformation. Finally, as was shown recently$^{13}$,
the spectral parameter arises as the constant valued
statistical gauge field of compactified 2+1 dimensional theory.
All results mentioned above suggest that the spectral parameter
is deeply connected with Abelian gauge field.
\par
Let us consider the moving frame equations (3.16),
$$
D_{0}{\bf n}_{+} = - 2giD_{1}\psi {\bf s}
+ 2ig|\psi|^{2}{\bf n}_{+},\,\,
D_{1}{\bf n}_{+} = - 2g\psi{\bf s} ,\eqno (3.39a,b)
$$
$$\partial _{0}{\bf s} = iD_{1}\psi {\bf n}_{-} - i\bar D_{1}\bar \psi
{\bf n}_{+}\,\,,\,\,
\partial _{1}{\bf s} = \psi {\bf n}_{-} + \bar \psi {\bf n}_{+}
,\eqno (3.39c,d)$$
written in terms of the zero strength gauge potential $W_{\mu}$,
(2.13),
$$\partial_{0} W_{1} - \partial_{1} W_{0} = 0. \eqno(3.39e)$$
The general solution of the last equation is of the pure gauge
form given in terms of the real function,
$$W_{\mu} = \partial_{\mu} \alpha. \eqno(3.40)$$
For a restricted class of functions, linear
in $x$ and $t$, we put
$$W_{1} = 2\lambda = const. \eqno(3.41)$$
In this case the NLS (2.7a) becomes,
$$
i\partial_{0}\psi - 2i\lambda\partial_{1}\psi
+ \partial^{2}_{1}\psi + 2g|\psi|^{2}\psi = 0, \eqno(3.42)$$
where we have to choose $W_{0} = 1/2 W^{2}_{1} = 2\lambda^{2}$.
In the constant velocity frame , $v = - 2\lambda$,  this equation
acquires the normal form (2.7a),
$$
i\partial_{0'}\psi
+ \partial^{2}_{1'}\psi + 2g|\psi|^{2}\psi = 0, \eqno(3.43)$$
and any $\lambda$ dependence  disappears.
However, it remains in the corresponding
moving frame system (3.39) having the form,
$$
\partial_{1'}\left(\matrix{{\bf n}_{+} \cr {\bf n}_{-} \cr
{\bf s} \cr} \right) =
\left(\matrix{
i\lambda & 0 & -2g\psi \cr
0 & - i\lambda & -2g\bar\psi \cr
\bar\psi & \psi & 0 \cr}\right)
\left(\matrix{{\bf n}_{+} \cr {\bf n}_{-} \cr
{\bf s} \cr} \right),
\eqno(3.44a)$$
$$
\partial_{0'}\left(\matrix{{\bf n}_{+} \cr {\bf n}_{-} \cr
{\bf s} \cr} \right) =
\left(\matrix{
- i\lambda^{2} + 2ig|\psi|^{2} & 0 & -2ig\partial_{1'}\psi +
2g\lambda\psi \cr
0 & i\lambda^{2} - 2ig|\psi|^{2} & 2ig\partial_{1'}\bar\psi
+ 2g\lambda\bar\psi \cr
-(i\partial_{1'}\bar\psi + \lambda\bar\psi) &
i\partial_{1'}\psi - \lambda\psi & 0 \cr}\right)
\left(\matrix{{\bf n}_{+} \cr {\bf n}_{-} \cr
{\bf s} \cr} \right).
\eqno(3.44b)$$
In the normal basis ${\bf n}_{1},{\bf n}_{2},{\bf s}$, we have
equations
$$
\partial_{1'}\left(\matrix{{\bf n}_{1} \cr {\bf n}_{2} \cr
{\bf s} \cr} \right) =
\left(\matrix{
0 & -\lambda & -gU_{1} \cr
\lambda & 0  & -gU_{2} \cr
U_{1} & U_{2} & 0 \cr}\right)
\left(\matrix{{\bf n}_{1} \cr {\bf n}_{2} \cr
{\bf s} \cr} \right)
\eqno(3.45a)$$
$$
\partial_{0'}\left(\matrix{{\bf n}_{1} \cr {\bf n}_{2} \cr
{\bf s} \cr} \right) =
\left(\matrix{ 0 &
\lambda^{2} - 2g|\psi|^{2} & g\partial_{1'}U_{2}
+ g\lambda U_{1} \cr
- \lambda^{2} + 2g|\psi|^{2} & 0 & -g\partial_{1'}U_{1}
+ g\lambda U_{2} \cr
-\partial_{1'}U_{2} -  \lambda U_{1} &
\partial_{1'} U_{1} - \lambda U_{2} & 0 \cr}\right)
\left(\matrix{{\bf n}_{1} \cr {\bf n}_{2} \cr
{\bf s} \cr} \right),\eqno(3.45b)$$
where we defined (see eqs.(3.10))
$U_{1} = 2 \Re (\psi), U_{2} = 2 \Im (\psi)$.
The first system (3.45a) in terms of the gauge invariant fields
as in (2.12),($\lambda = 0$),
has the form of the Frenet equations for the
curve. This suggests the idea to interpret the gauge invariant
form of (3.45) as a moving curve equations$^{19}$.
But due to the essential role of the space-time curvature
which is proportional to the nonlinearity and can be negative
this analogy doesn't seem so useful. In opposite, we find that a
more general setting - the moving frame method is going deeply in to the
problem and naturally illuminates the gauge theoretical
structure of it.
\par Note that the potentials choice,
$$W_{1}  = 2\lambda, W_{0} = 2\lambda^{2},\eqno(3.46)$$
which allowed us to remove the $\lambda$ dependence from equations
of motion (3.42) is exactly the 1-cocycle (3.35b).
To reproduce the usual Lax type representation we use the
spinor representation of system (3.45),
and  according (3.6) we have $2 \times 2$
Zakharov-Shabat spectral problem
$$
J_{1} = + {i \over 2}\lambda \sigma_{3} + \left(
\matrix{0  & -g\bar\psi  \cr
\psi  & 0 \cr }\right),
\eqno (3.47a)$$
$$
J_{0} = i\sigma_{3}[-{\lambda^{2} \over 2} + g|\psi|^{2}] +
\left( \matrix{ 0 & g(i\partial_{1} + \lambda)\bar\psi   \cr
(i\partial_{1} - \lambda)\psi    &    0  \cr  }\right).
\eqno (3.47b)$$
for the NLS (3.43).
\par  To derive the hierarchy of moving frames, providing the linear
problem for any of equations (2.35) we need to modify the above
Galileo transformation approach (3.42-43). Let us first rewrite the
hierarchy (2.30) for the constant potentials $W_{\mu} $, denoting
 $W_{1} = 2\lambda$:
$$
i\partial_{0_{n}}\left(\matrix{ \psi \cr
- \bar\psi \cr} \right)
+ {1\over 2}W_{0_{n}}
\left(\matrix{ \psi \cr \bar\psi \cr} \right)
= ({\cal L} + \lambda I)^{n}
\left(\matrix{ \psi\cr \bar\psi \cr} \right) =
\sum^{n}_{k = 0} \left(\matrix{k \cr n \cr }\right)
\lambda^{n - k}{\cal L}^{k}
\left(\matrix{ \psi \cr \bar\psi \cr} \right) =
$$
$$
\sum^{n}_{k = 0} \left(\matrix{k \cr n \cr }\right)
\lambda^{n - k}
\left(\matrix{ \psi^{(k)}_{0} \cr \bar\psi^{(k)}_{0} \cr} \right) =
\lambda^{n}
\left(\matrix{ \psi\cr \bar\psi \cr} \right) +
\sum^{n - 1}_{k = 1} \left(\matrix{k \cr n \cr }\right)
\lambda^{n - k}
\left(\matrix{ \psi^{(k)}_{0} \cr \bar\psi^{(k)}_{0} \cr} \right)
+
\left(\matrix{ \psi^{(n)}_{0} \cr \bar\psi^{(n)}_{0} \cr} \right) =
$$
$$
\lambda^{n}
\left(\matrix{ \psi\cr \bar\psi \cr} \right) +
\sum^{n - 1}_{k = 1} \left(\matrix{k \cr n \cr }\right)
\lambda^{n - k}i\sigma_{3}\partial_{0_{k}}
\left(\matrix{ \psi \cr \bar\psi \cr} \right)
+  {\cal L}^{n}
\left(\matrix{ \psi \cr \bar\psi \cr} \right),
\eqno(3.48)$$
where $[{\cal L},\lambda I] = 0$ is used.
Choosing the gauge potentials $W_{0_{n}} = 2\lambda^{n}$, and
collecting all terms with the time hierarchy derivatives in the
left hand side  we get,
$$i\{ \partial_{0_{n}} -
\sum^{n - 1}_{k = 1} \left(\matrix{k \cr n \cr }\right)
\lambda^{n - k}\partial_{0_{k}} \}
\left(\matrix{ \psi \cr -\bar\psi \cr} \right) =
{\cal L}^{n}
\left(\matrix{ \psi \cr \bar\psi \cr} \right). \eqno(3.49)$$
Defining the new time hierarchy $\{t_{0'_{1}},
t_{0'_{2}}, ..., t_{0'_{n}},...\}$, such that,
$$
\partial_{0'_{n}} = \partial_{0_{n}} -
\sum^{n - 1}_{k = 1} \left(\matrix{k \cr n \cr }\right)
\lambda^{n - k}\partial_{0_{k}},\eqno(3.50)$$
we obtain for (3.49) the usual NLS hierarchy form (2.35):
$$i\sigma_{3}\partial_{0'_{n}}
\left(\matrix{ \psi \cr \bar\psi \cr} \right) =
{\cal L}^{n}
\left(\matrix{ \psi \cr \bar\psi \cr} \right). \eqno(3.51)$$
The matrix structure of the above time hierarchy transformation
can be realized by the upper triangular matrix
$$\left(\matrix{ t'_{1} \cr t'_{2} \cr t'_{3} \cr : \cr t'_{n}}\right)
= \left(\matrix{ 1 & a_{12} & a_{13} & .. & a_{1n}  \cr
                 0 & 1      & a_{23} & .. & a_{2n} \cr
                 0 & 0      & 1      & .. & a_{3n} \cr
                 0 & :      & 0      & .. & :      \cr
                 0 & 0      & 0      & .. & 1      \cr } \right)
\left(\matrix{ t_{1} \cr t_{2} \cr t_{3} \cr : \cr t_{n} } \right)
,\eqno(3.52)$$
while the time derivatives transform by the lower triangular matrix
$$\left(\matrix{ \partial_{0_{1}} \cr \partial_{0_{2}} \cr
\partial_{0_{3}} \cr : \cr \partial_{0_{n}}}\right)
= \left(\matrix{1 & 0      & 0      & .. & 0 \cr
           a_{12} & 1      & 0      & .. & 0 \cr
           a_{13} & a_{23} & 1      & .. & 0 \cr
               :  & :      &      : & .. & 0 \cr
           a_{1n} & a_{2n} & a_{3n} & .. & 1 \cr }\right)
\left(\matrix{ \partial_{0'_{1}} \cr \partial_{0'_{2}} \cr
\partial_{0'_{3}} \cr : \cr \partial_{0'_{n}}}\right), \eqno(3.53)$$
with coefficients chosen as an appropriate  polynomial
of the parameter $\lambda$.
\par
By using the moving frame formulas (3.21-24) we can construct the
hierarchy of moving frames, which provides also the linear problem for
any member of the hierarchy (3.51).
For the space part we have,
$$
\left(\matrix{D_{1}{\bf n}_{+} \cr \bar D_{1}{\bf n}_{-} \cr
\partial_{1}{\bf s} \cr} \right) =
\left(\matrix{
0 & 0 & -2g\psi \cr
0 & 0 & -2g\bar\psi \cr
\bar\psi & \psi & 0 \cr}\right)
\left(\matrix{{\bf n}_{+} \cr {\bf n}_{-} \cr
{\bf s} \cr} \right),
\eqno(3.54)$$
in which $W_{1} = 2\lambda$ gives us the space part of the
moving frame system and is universal for
the whole hierarchy
$$
\partial_{1}\left(\matrix{{\bf n}_{+} \cr {\bf n}_{-} \cr
{\bf s} \cr} \right) =
\left(\matrix{
i\lambda & 0 & -2g\psi \cr
0 & - i\lambda & -2g\bar\psi \cr
\bar\psi & \psi & 0 \cr}\right)
\left(\matrix{{\bf n}_{+} \cr {\bf n}_{-} \cr
{\bf s} \cr} \right)
\eqno(3.53)$$
For the time evolution we start from the system for the
$t_{n}$ time:
$$
\left(\matrix{D_{0_{n}}{\bf n}_{+} \cr
\bar D_{0_{n}}{\bf n}_{-} \cr
\partial_{0_{n}}{\bf s} \cr} \right) = -2g
\left(\matrix{
{\bf s} - {\bf n}_{+}\int^{x}\bar \psi & {\bf n}_{+}\int^{x}\psi \cr
{\bf n}_{-}\int^{x} \bar\psi & {\bf s} - {\bf n}_{-}\int^{x}\psi \cr
-{1\over 2g}{\bf n}_{-} & -{1\over 2g}{\bf n}_{+} \cr}\right)
\left(\matrix{ \psi^{(n-1)}_{0} \cr \bar\psi^{(n-1)}_{0} \cr } \right).
\eqno(3.54)$$
After substituting the value $W_{0_{n}} = 2\lambda^{n}$ and
(2.25), we have the expression
$$
\partial_{0_{n}}\left(\matrix{{\bf n}_{+} \cr {\bf n}_{-} \cr
{\bf s} \cr} \right) = i\lambda^{n}
\left(\matrix{ {\bf n}_{+} \cr -{\bf n}_{-}\cr 0 \cr }     \right)
-2g
\left(\matrix{
{\bf s} - {\bf n}_{+}\int^{x}\bar \psi & {\bf n}_{+}\int^{x}\psi \cr
{\bf n}_{-}\int^{x} \bar\psi & {\bf s} - {\bf n}_{-}\int^{x}\psi \cr
-{1\over 2g}{\bf n}_{-} & -{1\over 2g}{\bf n}_{+} \cr}\right)
\Lambda^{n - 1}
\left(\matrix{ \psi \cr \bar\psi \cr } \right).
\eqno(3.55)$$
Then, for the transformed time hierarchy using
(3.50) and (3.55), we get
$$
\partial_{0'_{n}}\left(\matrix{{\bf n}_{+} \cr {\bf n}_{-} \cr
{\bf s} \cr} \right) = i\lambda^{n}(3 - 2^{n})
\left(\matrix{1 & 0 & 0 \cr
              0 & -1 & 0 \cr
             0 & 0 & 0 \cr }\right)
\left(\matrix{ {\bf n}_{+} \cr {\bf n}_{-}\cr {\bf s} \cr }  \right) -
$$
$$2g
\left(\matrix{
{\bf s} - {\bf n}_{+}\int^{x}\bar \psi & {\bf n}_{+}\int^{x}\psi \cr
{\bf n}_{-}\int^{x} \bar\psi & {\bf s} - {\bf n}_{-}\int^{x}\psi \cr
-{1\over 2g}{\bf n}_{-} & -{1\over 2g}{\bf n}_{+} \cr}\right)
\{\Lambda^{n - 1} - \sum^{n - 1}_{k = 1} \left(\matrix{ k \cr n \cr}
\right)\lambda^{n - k}\Lambda^{k - 1}\}
\left(\matrix{ \psi \cr \bar\psi \cr } \right)
\eqno(3.56)$$
The system (3.53), (3.56) is the moving frame system for the
hierarchy (3.51). The space part is linear in the spectral
parameter
$\lambda$, while the time part contains the $n$-th degree polynomial
for the $n$-th member of the hierarchy. To reproduce this
dependence in the explicit
form, according to (2.23a) ,
we decompose the gauge covariant operator
$\Lambda = {\cal L} + 2 I$. As a result we have the moving frame
evolution in terms of the usual AKNS operator ${\cal L}$,
$$
\partial_{0'_{n}}\left(\matrix{{\bf n}_{+} \cr {\bf n}_{-} \cr
{\bf s} \cr} \right) = i\lambda^{n}(3 - 2^{n})
\left(\matrix{1 & 0 & 0 \cr
              0 & -1 & 0 \cr
             0 & 0 & 0 \cr }\right)
\left(\matrix{ {\bf n}_{+} \cr {\bf n}_{-}\cr {\bf s} \cr }
\right) - $$
$$2g
\left(\matrix{
{\bf s} - {\bf n}_{+}\int^{x}\bar \psi & {\bf n}_{+}\int^{x}\psi \cr
{\bf n}_{-}\int^{x} \bar\psi & {\bf s} - {\bf n}_{-}\int^{x}\psi \cr
-{1\over 2g}{\bf n}_{-} & -{1\over 2g}{\bf n}_{+} \cr}\right)
\{\sum^{n - 1}_{l = 0} \left(\matrix{l \cr n - 1 \cr } \right)
 - $$
$$\sum^{n - 1}_{k = 1} \sum^{k - 1}_{l = 0}
\left(\matrix{ k \cr n \cr} \right) \left(\matrix{l \cr k - 1 \cr}
\right)\} \lambda^{n - 1 - l}{\cal L}^{l}
\left(\matrix{ \psi \cr \bar\psi \cr } \right)
\eqno(3.57)$$
For $n = 2$ these moving frame system coincides with the one (3.44)
obtained by the Galileo transformations. This suggests one to
construct the hierarchy of the Galileo transformations
with the properly defined time hierarchy for any of equations (3.51)
and the 1-cocycle conditions.
\bigskip
\noindent
{\bf IV. BF GAUGE THEORY AND THE TIME HIERARCHY}
\bigskip\par
In the present section we map the integrable
hierarchy from Sec.III to the non-Abelian BF gauge theory.
The BF theory is described by the action
$$S = {kL\over 4\pi}\int_{\Sigma_{2}}Tr(BF), \eqno(4.1)$$
where, $F = \epsilon^{\mu\nu}F_{\mu\nu}$,
is the gauge curvature tensor
in the adjoint representation of the appropriately
chosen group, while
$B$ is the world scalar Lagrange multipliers,
transforming according to the coadjoint representation$^{11}$.
This model, formulated in terms of 2-form $F$,
and the zero-form $B$ for any
2-dimensional manifold $\Sigma_{2}$ defines the topological field
theory$^{20}$. For the non-compact groups $SO(2,1)$
and $ISO(1,1)$
it provides a gauge theoretical formulation of the
Jackiw-Teitelboim lineal gravity
theories$^{11,21,22}$. This model,
being the constant curvature theory, includes the world scalar
Lagrangian multiplier field together with the Riemann scalar $R$
and is a type of scalar-tensor theories or the "dilaton"
gravity.
In the gauge formulation of the gravity theory  one introduces
instead
of the metric tensor $g_{\mu\nu}$, the Einstein-Cartan zweibeins
and the spin-connection
viewed as independent variables. Then,
the diffeomorphism-invariance of the gravity
theory becomes a part of
the local gauge invariance. The matter $ \psi $ and the gauge field
$V$ from the Sec.II are just the zweibeins and the spin-connection
fields$^{23}$, defining the moving frames equations.  The
$SU(2)$ and $SU(1,1)$  with the compact
Abelian $U(1)$ subgroup
in our case, correspond to the Euclidean gravity, with the local
$O(2)$ rotations in the tangent plane,
insted of the $O(1,1)$ Lorentz rotations.
\par
To describe the time evolution we suppose that $\Sigma_{2} =
R \times T$. Then the action (4.1) is,
$$S = {kL\over 2\pi}\int Tr B(\partial_{0}J_{1} -
\partial_{1}J_{0} + [J_{0},J_{1}])d^{2}x =
{kL\over 2\pi}\int Tr (B\partial_{0}J_{1} + J_{0}(\partial_{1}B +
[J_{1},B]))d^{2}x,\eqno(4.2)$$
where we integrated the second term by parts.
In terms of (3.8) parametrization
and with $B$ field given by
$B = i\phi_{0}\sigma_{3} + g\phi\sigma_{-} - \bar\phi\sigma_{+}$
the action becomes,
$$S = \gamma \int (\bar\phi\partial_{0}\psi +
\phi\partial_{0}\bar\psi
 + {1\over 2}\phi_{0}\partial_{0}V_{1}
 - {\cal H})d^{2}x,\eqno(4.3)$$
where the Hamiltonian density is given by,
$${\cal H} = -\{ \bar\psi_{0}(D_{1}\phi + 2ig\phi_{0}\psi)+
\psi_{0}(\bar D_{1}\bar\phi - 2ig\phi_{0}\bar\psi)
              + V_{0}{1\over 2}(\partial_{1}\phi_{0}
+ i\phi\bar\psi
- i\bar\phi\psi)\},
\eqno(4.4)$$
and $\gamma$ is the renormalized coupling constant.
Since the action is already in the first order form we can
immediately
write the canonical brackets$^{24}$:
$$
\{\psi(x),\bar\phi(y)\} = {1\over \gamma}\delta (x - y),
\,\,
\{\bar\psi(x),\phi(y)\} = {1\over \gamma}\delta (x - y),$$
$$ \{V_{1}(x),\phi_{0}(y)\} = {2\over \gamma}\delta (x - y),
\eqno(4.5)$$
The nondynamical Lagrange multipliers $\psi_{0}$,$\bar\psi_{0}$
and $V_{0}$ lead to the BF "Gauss-law" constraints
$$ D_{1}\phi + 2ig\psi\phi_{0} = 0,\,
\bar D_{1}\bar\phi - 2ig\bar\psi\phi_{0} = 0,\eqno(4.6)$$
$$ \partial_{1}\phi_{0} + i\phi\bar\psi - i\bar\phi\psi = 0,
\eqno(4.7)$$
Due to the canonical brackets (4.5) the Gauss's law
constraints generate
the $SU(2)$ (g = 1), or $SU(1,1)$ (g = -1), algebra of
the gauge transformations in the dynamical fields
$\psi$,$\bar\psi$
and $V_{1}$. Since Hamiltonian (4.4) is proportional
to the constraints
(4.6),(4.7) it is weakly vanishing $H \approx 0$. The above
properties characterise  any reparametrization invariant theory.
\par
We can partially solve constraints (4.6),(4.7) by integration of
the last one
$$\phi_{0} = -i\int^{x}(\bar\psi\phi  -  \psi\bar\phi)\eqno(4.8)$$
Then, substituting to the first two (4.6), we get
$${\cal M}_{\psi}(\phi) \equiv D_{1}\phi + 2g\psi\int^{x}
(\bar\psi\phi - \psi\bar\phi)(y)dy = 0,\eqno(4.9a)$$
$${\overline {{\cal M}_{\psi}(\phi)} } \equiv \bar D_{1}\bar\phi
- 2g\bar\psi\int^{x}
(\bar\psi\phi - \psi\bar\phi)(y)dy = 0,\eqno(4.9b)$$
where we defined the integro-differential operators ${\cal M}$.
Combined these two relations can be written in terms of
U(1) gauge covariant $\Lambda$ operator (2.22):
$$i\sigma_{3}\left(\matrix{{\cal M}_{\psi}(\phi) \cr
   {\overline {{\cal M}_{\psi}(\phi)}} \cr}\right)
= \Lambda \left( \matrix{\phi \cr \bar\phi \cr  } \right)
= 0 \eqno(4.10)$$
This means that the constraint surface of the BF theory is defined
by zero modes of the covariant $\Lambda$ operator. In terms
of the usual ${\cal L}$ operator (2.23) we have the relation
$${\cal L} \left( \matrix{\phi \cr \bar\phi \cr  }  \right)
= - {1\over 2}V_{1} \left( \matrix{\phi \cr \bar\phi \cr  }\right)
, \eqno(4.11)$$
which for the constant valued gauge potential, like in (3.36),
$V_{1} = -2\lambda$, becomes the eigenvalue problem for the
operator ${\cal L}$:
$${\cal L} \left( \matrix{\phi \cr \bar\phi \cr  }  \right)
= \lambda \left( \matrix{\phi \cr \bar\phi \cr  }\right)
 \eqno(4.12)$$
Comparing with the theory of solitons$^{12}$ show that our
world scalars $\phi$ and $\bar\phi$ are very similar to the squared
eigenfunctions for the Zhakharov-Shabat problem. Moreover,
the constraints surface system (4.6),(4.7) is related
to the Maxwell-Bloch equations$^{25}$, describing the
propagation of pulses through an inhomogeneously
broadened and resonant medium, where
$\psi$ is the electric field envelope,
$\phi$ is the polarization
and $\phi_{0}$ is the excitation number density.
It would be interesting to illuminate the relation between
last problem and the BF theory.
\par
We can express the action (4.3) in terms of the bilinear form.
For any two complex functions $\phi$ and $\chi$ vanishing at
infinities we define:
$$<\phi,\chi> = \int^{\infty}_{-\infty}(\bar\phi\chi +
\phi\bar\chi)
dx = 2 \Re \int^{\infty}_{-\infty} \bar\phi \chi dx,
\eqno(4.13)$$
Then, the operator ${\cal M}_{\psi}(\phi)$ from (4.9)
is skew symmetric with
respect to this bilinear form
$$<\phi,{\cal M}\chi> = - <{\cal M}\phi, \chi> \eqno(4.14)$$
if for the integral part of  ${\cal M}$ we use the proper definition
(2.22a).
Another skew symmetric operator is the immaginary unit
$i = \sqrt {-1}$,
$$<\phi, i\chi> = - <i\phi, \chi> \eqno(4.15)$$
In spite of the skew symmetry of these two operators, the
product
$${\cal R} = i{\cal M}, \eqno(4.16)$$
being the recursion operator, is not hermitian. This is due to the
noncommutativity of the operators,
$$[i,{\cal M}](\phi) = - 4ig\psi\int^{x} \bar \phi \psi
\eqno(4.17)$$
\par To derive the Hamiltonian dynamics for action (4.3) we
introduce the $U(1)$ gauge invariant variables.
Representing potential $V_{1}$ in a pure gauge form,
$$V_{1} = 2\partial_{1}\alpha(x), \eqno(4.18)$$
for the gauge invariant B(x)
(the magnetic field analog),
$$\phi_{0}(x) = \int^{x} B(y)dy, \eqno(4.19)$$
the Poisson bracket (4.5) results from
the following
canonical bracket,
$$\{\alpha(x),B(y)\} = - {1\over \gamma}\delta (x - y).
\eqno(4.20)$$
Defining the $U(1)$ gauge invariant fields,
$$\psi = \Psi e^{i\alpha},\,\, \phi = \Phi e^{i\alpha},
\eqno(4.21)$$
for action (4.3) we get,
$$S = \gamma\int [\bar\Phi\partial_{0}\Psi +
\Phi\partial_{0}\bar\Psi
+ \{\bar\psi_{0} e^{i\alpha}(\partial_{1}\Phi + 2ig\Psi \int^{x}B)
+ c.c.   \} +  $$
$${1 \over 2}(V_{0} - 2\partial_{0}\alpha)
\{B + i\bar\Psi\Phi - i\bar\Phi\Psi \}]. \eqno(4.22) $$
After redefinition of the Lagrange multipliers,
$${\cal V}_{0} \equiv V_{0} - 2\partial_{0}\alpha,\,\,
\bar\Psi_{0} \equiv \bar\psi_{0} e^{i\alpha},\,\,
\Psi_{0} \equiv \psi_{0} e^{-i\alpha} \eqno(4.23)$$
the action is written only in terms of the gauge invariant
dynamical variables $\Psi,\,\bar\Psi,\,\Phi,\,\bar\Phi $:
$$S = \gamma\int [\bar\Phi\partial_{0}\Psi + \Phi\partial_{0}\bar\Psi
+ \{\bar\Psi_{0} (\partial_{1}\Phi + 2ig\Psi \int^{x}B)
+ c.c.   \} + {1 \over 2}{\cal V}_{0}
\{B + i\bar\Psi\Phi - i\bar\Phi\Psi \}]. \eqno(4.24)$$
\par We can solve the Gauss law constraint explicitly,
$$ B = -i(\bar\Psi\Phi - \Psi\bar\Phi).\eqno(4.25)$$
Then the action is,
$$S = \gamma\int (\bar\Phi\partial_{0}\Psi +
\Phi\partial_{0}\bar\Psi)
- H, \eqno(4.26a)$$
where the Hamiltonian in terms of the bilinear form (4.13)
becomes,
$$H = -\gamma<\Psi_{0}, M_{\Psi}\Phi> =
       \gamma<M_{\Psi}\Psi_{0},\Phi>
\eqno(4.26b)$$
Here, the skew symmetric operator $M$ is the noncovariant form of
the ${\cal M}$ operator (4.9):
$$M_{\Psi}(\Phi) \equiv \partial_{1}\Phi + 2g\Psi\int^{x}
(\bar\Psi\Phi - \Psi\bar\Phi)(y)dy = 0,\eqno(4.27)$$
and coincides with the one introduced by Magri$^{26}$.
The Poisson brackets for $U(1)$ gauge invariant dynamical
variables,
$$\{\Psi (x), \bar\Phi (y)\} = {1\over \gamma}\delta (x - y),\,\,
  \{\bar\Psi (x), \Phi (y) \} = {1 \over \gamma} \delta (x - y)
\eqno(4.28)$$
generate the Hamiltonian equations given by,
$$\partial_{0}\Psi = \{\Psi,H\} = {1 \over \gamma}{{\delta H} \over
\delta \bar\Phi} = M_{\Psi}(\Psi_{0}),\eqno(4.29a) $$
$$\partial_{0}\bar\Psi = \{\bar\Psi,H\} =
-{1 \over \gamma}{{\delta H} \over
\delta \Phi} = \overline {M_{\Psi}(\Psi_{0})}, \eqno(4.29b)$$
or in terms of the recursion operator, $R = iM$,
$$i\partial_{0}\Psi = R_{\Psi}(\Psi_{0}),\,\,\,
\,\, -i\partial_{0}\bar\Psi = \overline {R_{\Psi}(\Psi_{0})}
\eqno(4.30)$$
Thus, we can see the effect of introducing the $\Phi$ field,
as the canonical momentum conjugate to the dynamical field $\Psi$,
is to eliminate the nondynamical Lagrange
multipliers $\Psi_{0}$ $^{21}$.
Due to the arbitrariness of these multipliers, by the
recursion operator $R$,
we can construct the whole
hierarchy of the multipliers,
$$ \Psi^{(n)}_{0} = R_{\Psi}(\Psi^{(n-1)}_{0}) = \,...\, =
R^{n}_{\Psi}(\Psi^{(0)}_{0}), \eqno(4.31)$$
Then, the related actions,
$$S_{n+1} = \gamma\int (\bar\Phi\partial_{0_{n+1}}\Psi +
\Phi\partial_{0_{n+1}}\bar\Psi)
- H_{n+1}, \eqno(4.32a)$$
with the Hamiltonian,
$$H_{n+1} = -\gamma<\Psi^{(n)}_{0}, M_{\Psi}\Phi> =
       \gamma<M_{\Psi}\Psi^{(n)}_{0},\Phi>,
\eqno(4.32b)$$
define the evolution hierarchy
$$i\partial_{0_{n+1}}\Psi = R^{n + 1}_{\Psi}(\Psi^{(0)}_{0}),
\,\,
-i\partial_{0_{n+1}}\bar\Psi =
\overline {R^{n + 1}_{\Psi}(\Psi^{(0)}_{0})}.
\eqno(4.33)$$
By skew symmetry property of the $M$ operator we have the relations,
$$H_{n+1} = -\gamma<R^{(n)}(\Psi^{(0)}_{0}), M(\Phi)> \, =
 -\gamma<R^{(n-1)}(\Psi^{(0)}_{0}), MR(\Phi)> \, = $$ $$
 -\gamma<R^{(n-2)}(\Psi^{(0)}_{0}), MR^{2}(\Phi)> \, = ...
 = -\gamma<\Psi^{(0)}_{0}, MR^{n}(\Phi)> \eqno(4.34)$$
or,
$$H_{n+1} = \gamma<iR^{(n+1)}(\Psi^{(0)}_{0}), \Phi> \, =
 \gamma<i\Psi^{(n+1)}_{0}, \Phi^{(0)}> \, =   $$ $$
 \gamma<\Psi^{(0)}_{0}, iR^{n+1}(\Phi)> \, =
 \gamma<\Psi^{(0)}_{0}, i\Phi^{(n+1)}>, \eqno(4.35)$$
where,
$$\Phi^{(n+1)} = R^{n+1}(\Phi^{(0)}),\,\, \Phi^{(0)}
\equiv \Phi. \eqno(4.36)$$
If the constraint surface is defined by the relation
$$R(\Phi^{(0)}) = 0, \eqno(4.37)$$
then it satisfies to the constraints,
$$R^{2}(\Phi^{(0)}) \, = R^{3}(\Phi^{(0)}) \, = \,...\,
= 0,  \eqno(4.38)$$
for any degree $n$. Moreover, the surface (4.37)
defines evolution,
$$i\partial_{0_{1}} \Psi = R(\Psi^{0}), \eqno(4.39)$$
while the higher degree surface (4.38)
generates the higher evolution,
$$i\partial_{0_{n}} \Psi = R^{n}(\Psi^{0}). \eqno(4.40)$$
This just confirms the evident fact , that the first member
of the hierarchy
(4.39) completely defines the higher evolutions.
\par
Hence, in the Hamiltonian interpretation,
the canonical bracket (4.28)
for the
first member of hierarchy (4.39):
$$\{\Psi (x), \bar\Phi^{(0)} (y)\} =
{1\over \gamma}\delta (x - y),\,\,
  \{\bar\Psi (x), \Phi^{(0)} (y) \} =
{1 \over \gamma} \delta (x - y),
\eqno(4.41)$$
defines the higher Poisson brackets
$$\{\Psi (x), \bar\Phi^{(n)} (y)\}
= \{\Psi (x), \overline {R^{n}\Phi^{(0)}}(y)\}
{1\over \gamma}\delta (x - y), $$
$$
  \{\bar\Psi (x), \Phi^{(n)} (y) \}
= \{\bar\Psi (x), R^{n}\Phi^{(0)}(y)\}
{1 \over \gamma} \delta (x - y).
\eqno(4.42)$$
In addition, every Hamiltonian (4.35) corresponds to the higher
integrals of motion for the model.
Thus, the BF formulation provides a natural description
of the whole integrable hierarchy.
\par The above procedure can be extenden to the case of
spectral parameter dependence, by reduction of the $U(1)$ connection
to the constant value, as in Sec.II and III.
The details of calculation would be published soon.
\vfill\eject
\noindent
{\bf V. CONCLUSION}
\bigskip\par
The main ingredient of our consideration is the
moving frame method. Historically $^{27}$
, the moving frame idea has origin
from mechanics where studying the rigid body motion,
one gets  a one-parameter
frames family depending on time and completely characterizing the
rigid motion. G.Darboux$^{28}$ and E.Cotton$^{29}$ has generalized
one-parameter frame to several parameters. The
richness of the theory was demonstrated by E. Cartan$^{30}$
({\it la m\'ethode du r\'ep\`ere mobile}) who successfully
applied it to research in differential geometry. We find that
this deep method is conceptually simple to study
integrable models and the underlying gauge theoretical structure.
\par
In the recent study of the dimensionally reduced Jackiw-Pi
model$^{4-6}$, to produce dynamics for the B field,
an extension of the
Lagrangian was given and the chiral solitons, propogating only in
one direction was constructed. From our consideration it is clear
that the chiral solitons can be created by the odd members of
the hierarchy with the odd dispersion. This fact is
valid even for the linear approximation. Thus, for the NLS (2.32)
the linear wave with the wave number $k$ is
propagating with velocity $v = k$, while the wave for $-k$ is
propagating in the opposite direction with the same velocity.
On the contrary, for the next member of the hierarchy, Eq.(2.33),
both waves with $+k$ and $-k$ are propogating in the same direction
with velocity $v = k^{2}$. The similar structure is valid
also for solitons. This shows that the chiral solitons in our
approach belong to
the odd members of the NLS hierarchy.
\par
In conclusion, as shown in the present paper, the 1+1 dimensional
projection of Chern-Simons gauge interacting theory contains
the hidden hierarchy of integrable systems. The statistical
gauge field, after projection, still plays the crucial role
defining spectral characteristics of the system.
The key point for the
construction is the mapping to non-Abelian BF theory,
which covers all the rich structure of integrable system $^{31-33}$.
The commutativity of spectral flows naturally provides
the hierarchy of the gauge fixing constraints and
the corresponding time
hierarchy.
We hope that our procedure could clarify some difficult questions
of the diffeomorphism invariance breaking
in the theories like the gravity, and the time
hierarchy creation with the physical degrees of freedom from
the reparametrization invariance.

\bigskip
\par
\noindent
{\bf ACKNOWLEDGMENTS}
\bigskip
\medskip
\par
The authors would like to thank S.S. Roan and S.S. Lin for stimulating
discussions.
One of the authors (O.K.P) would like to thank Prof. Fon-Che Liu
and
Institute of Mathematics,
Academia Sinica, Taipei,
for the warm hospitality.
This work was supported in part by the National Science
Council at Taipei,
Taiwan.
\medskip \par
\vfill\eject

\noindent
{\bf REFERENCES}
\medskip
\bigskip
\par \noindent
$^{1}$R. Jackiw and S. Y. Pi, Phys. Rev. Lett. {\bf 64}, 2969 (1990);
Phys. Rev. {\bf D42}, 3500 (1990).
\par\noindent
$^{2}$F. Wilczek, {\it Fractional Statistics and Anyon
Superconductivity} (World Scientific, Singapore, 1990).
\par\noindent
$^{3}${\it The Quantum Hall Effect}, eds. S. Girvin and R. Prange
(Springer-Verlag, 1990).
\par\noindent
$^{4}$S. J. Benetton Rabello, Phys. Rev. Lett. {\bf 76}, 4007 (1996).
\par\noindent
$^{5}$U. Aglietti, L. Griguolo, R. Jackiw, S.-Y. Pi and D. Seminara,
hep-th/9606141.
\par\noindent
$^{6}$R. Jackiw, A nonrelativistic chiral soliton in one dimension,
Preprint MIT - CTP N 2587, hep-th/9611185.
\par\noindent
$^{7}$L. Martina, O. K. Pashaev and G. Soliani,
Mod. Phys. Lett. {\bf A8}, 3241 (1993)
\par\noindent
$^{8}$O. K. Pashaev, Mod.Phys.Lett. {\bf A11}, 1713 (1996).
\par\noindent
$^{9}$V. E. Zakharov and A. B. Shabat, Sov. Phys. JETP, {\bf 34},
62 (1972).
\par\noindent
$^{10}$C. Teitelboim, Phys. Lett. {bf 126B}, 41 (1983), and in
{\it Quantum Theory of Gravity}, S. Christensen, ed. (A. Hilger, Bristol,
UK, 1984); R. Jackiw, in {\it Quantum Theory of Gravity},
S. Christensen, ed. (A. Hilger, Bristol, UK, 1984), Nucl. Phys.
{\bf B252}, 343 (1985).
\par\noindent
$^{11}$R. Jackiw, Teor. Math. Phys. {\bf 92}, 979 (1992).
\par\noindent
$^{12}$M. J. Ablowitz, D. J. Kaup, A. C. Newell and H. Segur,
Stud. Appl. Math. {\bf 53}, 249 (1974).
\par\noindent
$^{13}$O. K. Pashaev, J. Math. Phys. {\bf 37}, 4368 (1996).
\par\noindent
$^{14}$L. A. Takhtajan, Phys. Lett. {\bf A64}, 235 (1977).
\par\noindent
$^{15}$O. K. Pashaev and S. A. Sergeenkov, Physica {\bf A137},
282 (1986).
\par\noindent
$^{16}$R. Jackiw, {\it Diverse topics in theoretical and mathematical
physics}, (World Scientific, Singapore, 1995).
\par\noindent
$^{17}$M. Tinkham, {\it Introduction to Superconductivity},
(McGraw-Hill, 1975).
\par\noindent
$^{18}$V. G. Kadyshevsky, in Proc.
{\it "Problemi Kvantovoy Teorii Polya"
(Problems of Quantum Field Theory)}, Alushta 79, (Dubna, 1979).
\par\noindent
$^{19}$A. Doliwa and P. M. Santini, Phys. Lett. {\bf A185}, 373
(1994), and references therein.
\par\noindent
$^{20}$M. Blau and G. Thompson, Ann. Phys. {\bf 205}, 130 (1991);
E. D'Hoker, Phys. Lett. {\bf B264}, 101 (1991).
\par\noindent
$^{21}$K. Isler and C.A. Trugenberger, Phys. Rev. Lett. {\bf 63}
834 (1989).
\par\noindent
$^{22}$T. Fukuyama and K. Kamimura, Phys. Lett. {\bf B160}, 259
(1985); A. Chamseddine and D. Wyler, Phys. Lett. {\bf B228}, 75
(1989).
\par\noindent
$^{23}$L. Martina, O. K. Pashaev and G. Soliani, "Integrable
Dissipative Structures in the Gauge Theory of Gravity",
Preprint Lecce Univ. DFUL-1/02/97, (1997)
\par\noindent
$^{24}$L. Faddeev and R. Jackiw, Phys. Rev. Lett. {\bf 60}, 1692
(1988).
\par\noindent
$^{25}$M.J. Ablowitz, D.J.Kaup, and A.C. Newell,
J. Math. Phys. {\bf 15}, 1852
(1974).
\par\noindent
$^{26}$F. Magri, J. Math. Phys. {\bf 19}, 1156 (1978).
\par\noindent
$^{27}$S. S. Chern et al., {\it Lecture notes on differential
geometry}, (in Chinese, 1982).
\par\noindent
$^{28}$G. Darboux, {\it Lecons sur la theorie generale des surfaces},
(Paris, Gauthier-Villars;I (1887)).
\par\noindent
$^{29}$E. Cotton, {\it G\'en\'eralisation de
la th\'eorie du tri\'edre mobile},
Bull. Soc. Math., Paris, {\bf 33}, 42 (1905)
\par \noindent
$^{30}$E. Cartan, {\it Theorie des groupes
finis et continus et la geometrie
differentielle traitees par la methode du repere mobile},
(Paris,
Gauthier- Villars; (1937)).
\par\noindent
$^{31}$J.-H. Lee and O. K. Pashaev, Int. J. Mod. Phys. {\bf A12},
213 (1997).
\par\noindent
$^{32}$O. K. Pashaev,
"Integrable models as constrained topological gauge theory",
in Proc. of the II Conference on Constrained Dynamics and Quantum
Gravity, (in honor of Tullio Regge 65's), Santa Margerita, Italy,
1996, Nucl. Phys. Suppl. (in press).
\par\noindent
$^{33}$J.-H. Lee and O. K. Pashaev, Abelian gauge theory and
integrable $\sigma$ models, Preprint MIAS 97 - 3, 
Inst. of Math. , Academia Sinica , (Taipei), June (1997)
\end